\documentclass[prl,twocolumn,graphicx,amssymb,floatfix]{revtex4}

\usepackage{graphicx}
\begin{document}

\title{Tracing the past of a quantum particle }

\author{L. Vaidman}
\affiliation{ Raymond and Beverly Sackler School of Physics and Astronomy\\
 Tel-Aviv University, Tel-Aviv 69978, Israel}

\begin{abstract}
 The question: ``Where was a quantum particle between pre- and postselection measurements?'' is analyzed in view of a recent proposal that it was in the overlap of the forward and backward evolving wave functions. It is argued that this proposal corresponds not only to the criterion of where the particle leaves a weak trace, but also where the local interactions can affect the probability of postselection and where finding the particle in a strong nondemolition measurement is possible. The concept of a  ``secondary presence'' of a pre- and postselected particle where local interactions affect the weak trace in the overlap region is introduced.
     \end{abstract}
\maketitle

Recently I  proposed to describe the past of a   quantum particle between two measurements utilizing the weak trace it leaves \cite{V13}. The rational for this criterion is that all physical interactions are local.  According to my definition the particle was present in places where its coupling with some other objects  led to a nonvanishing effect on these objects.

Another approach to the question where the particle was, which also relies on the locality of interactions, is that the particle is present in all places where it  can be affected by other objects. In the standard formalism, a quantum particle is described by its forward evolving wave function. In the two-state vector formalism (TSVF) \cite{AV90}, which is more appropriate for analysing pre- and postselected quantum systems, it is also described by the backward evolving state. Thus, a possible definition  is that the particle was in every place where its forward or backward evolving wave functions do not vanish. Clearly, the wave functions of the particle can be affected by objects placed in any point where at least one of the waves is present. This is a consistent approach, but not very helpful.  It requires to keep records of too much information.

\begin{figure}[b]
  \includegraphics[width=7.4cm]{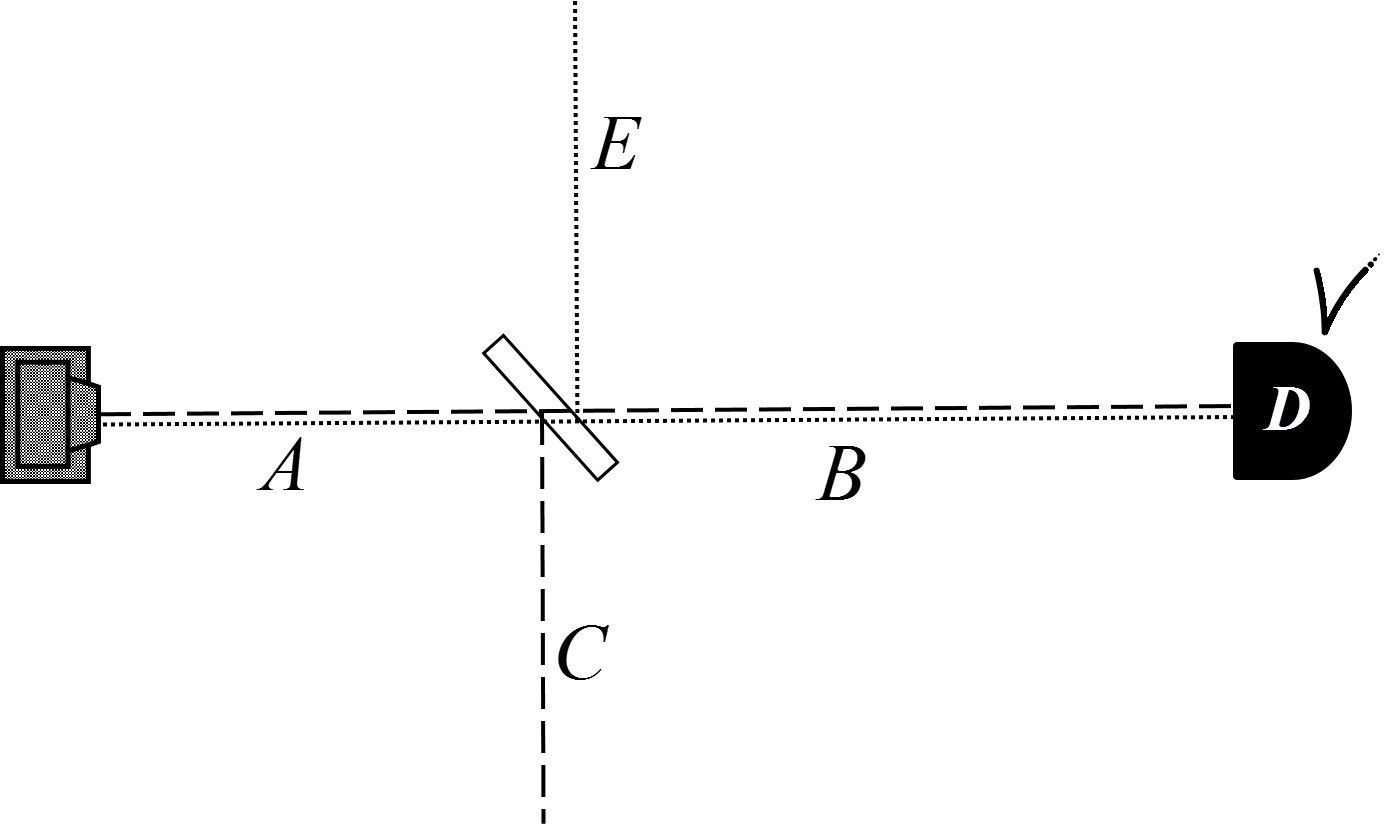}\\
     \caption{ A single particle  emitted by the source passes through a beam splitter and is absorbed by detector $D$. The forward (dashed line) and the backward (dotted line) evolving states are shown. Only in the overlap region of the forward  and backward  evolving states there are observable effects such as weak trace or a nonzero probability to find the particle in a nondemolition measurement. } \label{1}
\end{figure}

My guiding principle for describing the past of the particle is to consider  only observable aspects. Consider a simple experiment with a single photon source, a beam splitter and a detector shown on Fig. 1. An object, say an attenuator, placed in points $C$ or $E$, clearly changes the forward and backward evolving quantum states. However, when we restrict ourselves only to the cases in which the photon was observed by detector $D$, these changes are unobservable. The weak trace remains unchanged.   The probability of the postselection is not changed by local interactions in modes $C$ and $E$. In this simple setup  placing the attenuator anywhere in the overlap of the forward and backward wave function also does not change the weak trace that the pre- and postselected particle leaves, but the probability of the postselection does change. Thus,  in this case the criterion of the weak trace the particle leaves and the criterion of local interactions leading to an observable change on the particle provide the same picture of the past of the particle: the particle was in the overlap of the  forward and backward evolving quantum states. Note that this is also the place where the pre- and postselected particle can be found in a strong nondemolition measurement.

Consider now the example  which I suggested in \cite{V13}. It is a  Mach-Zehnder interferometer (MZI) nested inside another MZI, Fig. 2. Placing the attenuator at $E$ or at $D$, apart from changing the two-state vector describing the pre- and postselected particle,   has observable consequences: it affects the weak trace the particle leaves inside the inner interferometer. So, in some sense, the particle is present not only in the  overlap of the forward and backward evolving wave functions, but also  in the regions where the forward and backward evolving waves   lead towards the overlap. The  ``presence'' of the particle in this case is much weaker, henceforth I will refer to  it as a ``secondary presence''. The particle at the location of secondary presence does not affect other systems: it leaves no trace there and it cannot be found  there in a nondemolition measurement. It is also  not  affected by other systems as much as it is in the overlap region, still some effect on the particle exists. The interaction with other systems does not change the probability of the postselection. It does change, however, the weak trace the particle leaves in  the overlap region. It also changes the probability (conditional on postselection) to find the particle in a nondemolition measurement in that region.

\begin{figure}[t]
      \includegraphics[width=8cm]{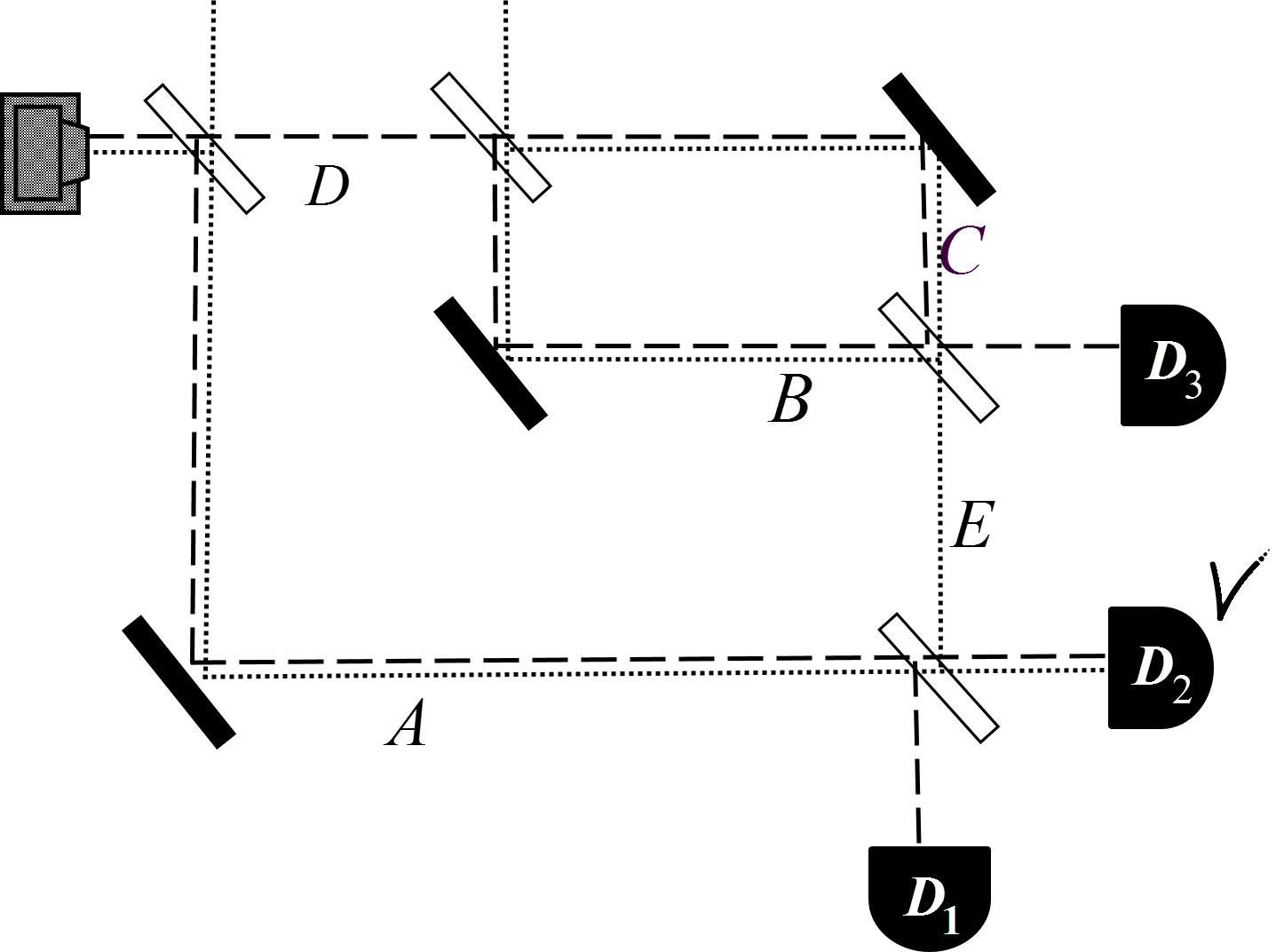}\\
   \caption{(Fig. 4c of \cite{V13}) The particle, detected by $D_2$ leaves a trace only where both forward (dashed line) and backward (dotted line) evolving states are present.  We suggest that there is a ``secondary presence'' of the particle in the segments $D$ and $E$. There is no weak trace there, the particle cannot be found there, but local interactions at these segments affect the weak trace inside the inner interferometer. } \label{2}
\end{figure}

The lack of reciprocity in the interaction between the particle and an object in the region of a ``secondary presence'' seems to contradict  physical intuition: local interaction between the particle and the object at $E$ leads to an observable change of the particle's weak trace while it changes nothing in the object. The key to this unusual behaviour is the postselection.
In the interpretation of quantum mechanics which I prefer, namely, the many-worlds interpretation (MWI) \cite{SEP}, the explanation is  as follows. Physical intuition is based on physical laws which are applicable to the physical Universe incorporating all the worlds. In our case, in addition to the world with the photon postselected at $D_2$, there are worlds with the particle detected at $D_1$, and $D_3$. In the  world of a supertechnology which can manipulate experimentalists in a superposition which consists of the worlds with all outcomes, no change of the weak trace of the particle takes place when  an object is placed in $E$.  We, in the world of postselection at $D_2$ (as ``we'' in the world of postselection at $D_1$) are in a privileged  position to see this (subjective) change.  In the physical universe incorporating all the worlds there is a reciprocity regarding the action of the particle on the object and the action of the object on the particle: in this experiment they do not affect each other.

When considering interactions of the particle with an object in $D$, reciprocity  is restored in another way. There is a world in which the particle is absorbed by the object. In this world and in the physical universe which incorporates all worlds together, the object is affected by the particle and the particle is affected by the object.

The time symmetry of the TSVF and the symmetry of the setup of Fig. 2. with respect to exchanging the source and the detector imply  that the secondary presence in  $D$ is the same as that in $E$. This, however is true only subjectively for the observer detecting the photon in $D_2$ ($D_1$). From the perspective of the physical universe all outcomes are realized, so the backward evolving quantum state includes wave packets ``emitted'' by all detectors. The backward evolving wave packets from $D_1$ and $D_2$ interfere  destructively in $E$. On the other hand, the backward evolving wave packet from $D_3$ reaches $D$ and no other wave packets interfere with it. So, from the point of view of the omnipotent technology capable of seeing all the worlds, there is no presence of the photon in $E$ of any kind and there is a ``full'' presence in $D$. There is a weak trace in $D$ and no weak trace in $E$. Thus, the ``secondary presence'' of a particle is a subjective concept of an observer who obtained a particular result in a postselection measurement. From the physical universe point of view, there is only primary presence and it is in every place where the forward evolving wave function does not vanish \cite{Vtime}.

\begin{figure}[b]
 \includegraphics[width=8cm]{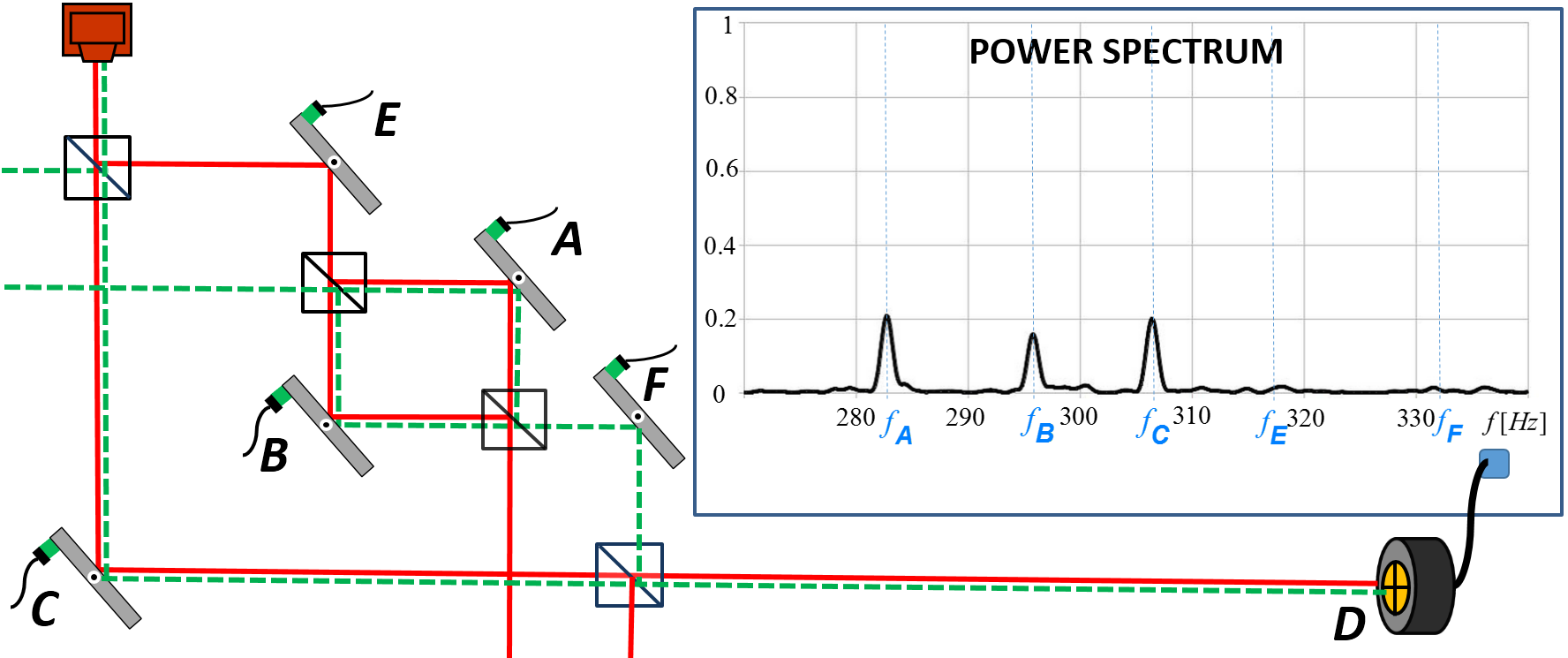}\\
      \caption{(Color online) (Fig. 3 of \cite{Danan})
      All mirrors perform tiny vibrations with different frequencies moving  reflected beams up and down.   A quad-cell photo-detector $D$ measures the signal, namely the difference of the currents generated in its upper and lower cells. The photons detected by $D$ are described by the forward evolving quantum state (red line) and  backward evolving quantum state (green dashed line). The frequencies $f_C, f_A$ and $f_B$ of vibration of the mirrors in the overlap region are observed while $f_E$ and  $f_F$ are not.  } \label{3}
\end{figure}

The main conceptual difference between the experiment demonstrating  the past of the photon \cite{Danan}, Fig. 3. and the setup of Fig. 2. is the fact that the weak trace is ``written'' on the photon itself. This made the measurement on a large postselected ensemble feasible. Considering the mode (the output port) as a complete postselection of the photon, while a small transversal deviation from the center of the beam in this mode measured by the quad-cell detector as the ``trace'' left by the particle, allows to view this experiment  as an implementation of the setup of Fig. 2. The weak trace was not measured at the location where it was created, but in the quad-cell detector. The locations of the creation of the trace were determined by the the frequencies of the disturbance due to vibrating mirrors. The picture according to which the photons were present at the overlap region of the forward and backward evolving states does account for the result: the trace was created only at the overlap region. This picture fits also another way to view this experiment: the photons bring the information about frequencies of vibrations of the mirrors they bounced off.

One might think that the above experiment tested the ``secondary presence'' of the photon at $E$ and $F$   and did not find it. The vibrating mirrors $E$ and $F$ have not led to an observable change. Blocking  $F$  eliminates the weak trace created at $A$ and $B$, see Fig. 4. The attenuator placed on the way to $E$ or $F$   reduces the peaks at $f_A$ and $f_B$. However, in the experiment described in Fig. 3, instead of attenuating the beam, only small disturbance was introduced in $E$ and $F$.  Tiny vibrations of the mirrors $E$ and $F$ cause only tiny modifications of the weak traces which are all small by themselves, so this second order effect has not been seen.

\begin{figure}[b]
  \includegraphics[width=8cm]{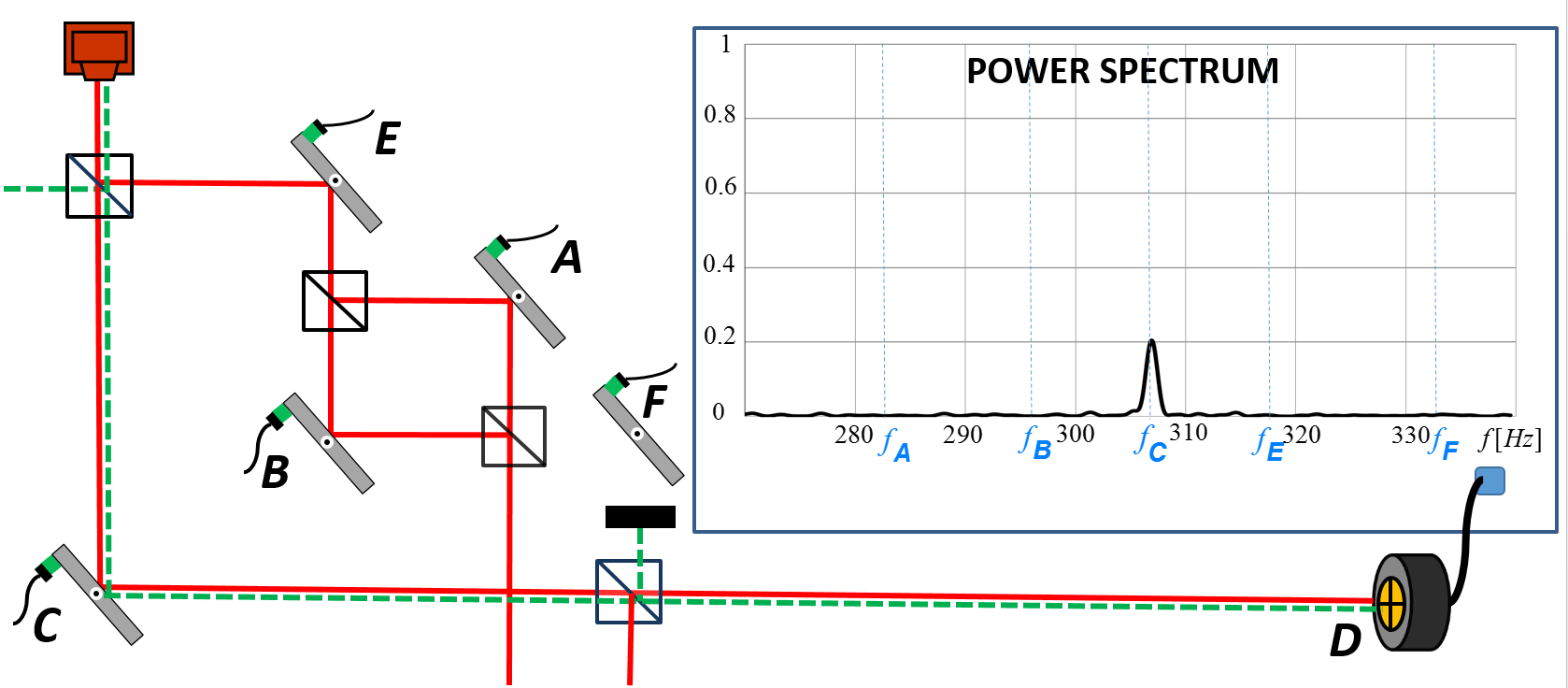}
  \caption{(Color online) (Fig. 5 of Supplement I of \cite{Danan})
  A block between mirror $F$ and the last beam splitter ``absorbs'' the backward evolving wave (dashed line) moving towards mirrors  $A$ and $B$, and eliminates the peaks  at $f_A$ and $f_B$.}
\end{figure}

The photons in the performed experiment were  not postselected in a completely specified  state. All photons reaching the quad-cell detector were considered, those detected by the upper part and those detected by the lower part of the detector.
The postselection was a partial measurement projecting not on a particular state, but on a particular subspace of states.
One approach to deal with a partially  postselected particle at the intermediate time is to wait until the future measurements will end up in a complete postselection which will specify the backward evolving state.  But it is a legitimate question to ask: ``What is the description of the particle  preselected in the state $|\Psi \rangle$    in the world of the partial postselection on the subspace $C$ which was not yet split by further measurements?" I propose to describe such a particle by the following two-state vector:
\begin{equation}\label{pps}
 \langle \Phi |~|\Psi\rangle, ~~~~~~~~  |\Phi\rangle \equiv \frac{{\rm \bf P}_C~|\Psi\rangle}{|{\rm \bf P}_C~|\Psi\rangle|}.\label{1}
\end{equation}
(An obvious modification due to the free Hamiltonian evolution is not shown here.)

 A simple argument supporting  this proposal  is that a projection measurement on the state $ |\Phi\rangle$ immediately after the partial postselection will succeed with certainty and then  $\langle \Phi |~|\Psi\rangle$ will be the description according to the standard TSVF approach. It is clear that we can apply this method only if no strong measurements have been performed at the intermediate time. Strong measurements  lead to the state  of the particle after the partial postselection to be  different from $|\Phi\rangle$, so the final projection measurement might not succeed. An example showing the failure of (\ref{1}) with intermediate strong measurements is the case of no postselection which corresponds to a projection on the whole space of states. Indeed, the probability, given by Aharonov, Bergmann and Lebowitz \cite{ABL}, of an outcome of a strong measurement performed on a particle described by the two-state vector $\langle \Psi |~|\Psi\rangle$  which corresponds to a special case of a partial postselection which is ``no postselection'', is different from the probability given by the standard quantum mechanics which is directly applicable for the case in which there is no postselection.

\begin{figure}[b]
  \includegraphics[width=8cm]{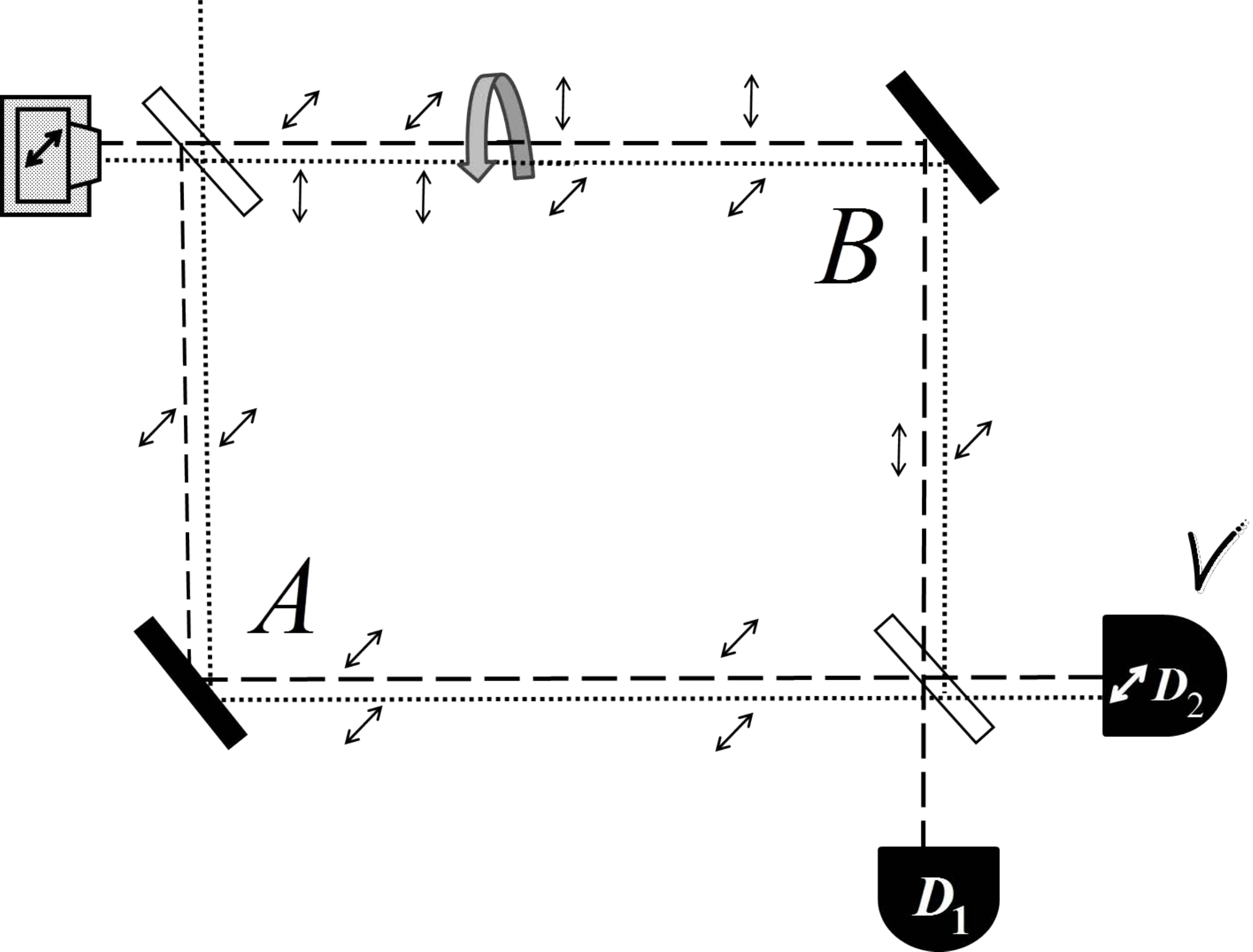}\\
    \caption{(Fig. 5 of \cite{V13}) The photon  in horizontally polarized state $|H\rangle$ enters the MZI with polarization-insensitive beam splitters.  The  polarization of the wave packet in arm $B$ is rotated to an orthogonal state $|V\rangle$. The photon is detected in $|H\rangle$ state. Aharonov et al. suggest that at $B$ there is only a ``grin'' of the photon, since polarization-insensitive nondemolition measurement  cannot find the photon in $B$.} \label{4}
\end{figure}

Let me note a semantic difference between my approach and the recent paper on Quantum Cheshire Cat  \cite{ChCat}.  Aharonov et al. considered a MZI with polarized photons in a  setup similar to what I considered in \cite{V13}, see Fig. 5. We all agree about the outcomes of measurements. Weak value $({\rm \bf P}_B )_w$ is zero. Thus, the outcome of the weak measurement of the presence of the photon in $B$, as well as the probability to find the photon there in a strong nondemolition polarization-insensitive measurement, vanish. In the rare event of obtaining a macroscopic ensemble of such pre- and postselected photons, all polarization-insensitive measurements  in $B$, such as energy flux, show null result, while polarization measurements yield a macroscopic effect. This led Aharonov et al. to write that the photon was not in $B$, only its polarization, ``the grin of Cheshire Cat'', was there.  The reason for this is that the weak value of the projection in $B$ on the circular polarization of the photon does not vanish, $({\rm \bf P}_B {\rm \bf P}_\circlearrowright )_w  ={1\over 2}$. Using the same equation and the observation that  local polarization-sensitive interactions exist, I chose to say  that the photon was in $B$. This is because the photon leaves a trace in $B$, there is a nonvanishing probability to find the photon with a particular polarization in  a nondemoltion measurement performed in $B$, and because placing a polarizer in $B$ changes the probability of the postselection of the photon.

To summarize, I suggest to say that a pre- and postselected quantum particle was in the region of the overlap of the forward and backward evolving wave functions. This is the case even when the forward and backward evolving waves differ in an internal degree of freedom, so long as these waves are not entangled with spatially separated systems. This proposal is supported by the following observable effects. In the overlap region and only in the overlap region:

i) The particle leaves a weak trace.

ii) There is a nonzero probability to find the particle in a nondemolition measurement. (Such a measurement might include projection on some internal degree of freedom.)

iii) Local interactions lead to a change in the probability of the postselection of the particle.

I  also define a ``secondary presence'' in the regions of the forward and backward evolving wave functions leading toward the overlap. Although (i)-(iii) do not hold there, local interactions in this region change the weak trace the particle leaves in the overlap region as well as the probability of finding the particle in a nondemolition measurement there.


I thank Eliahu Cohen, Paul-Antoine Moreau and Shmuel Nussinov for helpful discussions. This work has been supported in part by  grant number 32/08 of the Binational Science Foundation and the Israel Science Foundation  Grant No. 1125/10.


\begin{thebibliography}{99}

\bibitem{V13}
L. Vaidman,
Phys.  Rev. A  {\bf 87}, 052104 (2013).


\bibitem{AV90}
Y. Aharonov  and L. Vaidman,
   Phys. Rev. A \textbf{41}, 11 (1990).



\bibitem{Danan}
 A. Danan, D. Farfurnik, S. Bar-Ad and L. Vaidman,
Phys. Rev. Lett. {\bf 111}, 240402 (2013).

\bibitem{SEP} L.~Vaidman,  Many-Worlds Interpretation of Quantum
Mechanics, {\it Stan. Enc. Phil.},  E. N. Zalta (ed.) (2002),
http://plato.stanford.edu/entries/qm-manyworlds/.


\bibitem{Vtime}
L. Vaidman,
Time Symmetry and the Many-Worlds Interpretation,
in {\it Many Worlds? Everett, Quantum Theory, and Reality},
S. Saunders, J. Barrett, A. Kent, and D. Wallace eds., (Oxford University Press 2010).


\bibitem{ABL}
 Y. Aharonov, P. G. Bergmann,  and J. L.   Lebowitz,    Phys.  Rev.  {\bf B134}, 1410 (1964).

\bibitem{ChCat}
Y. Aharonov, S. Popescu, D. Rohrlich, and P. Skrzypczyk,
 New J. Phys. {\bf 15}, 113015 (2013).


\end{thebibliography}
\end{document}